\documentstyle[aaspp4,psfig]{article}
\begin{document}

%newcommand----------------------------------------------------------------
\def\be{\begin{equation}}           \def\ee{\end{equation}}
\def\lsim{\lower0.5ex\hbox{$\; \buildrel < \over \sim \;$}}
\def\gsim{\lower0.5ex\hbox{$\; \buildrel > \over \sim \;$}}
\def\ws{\mbox{$\omega_{\rm s}$}}
\def\ergs{\mbox{erg\,s$^{-1}$}}     
\def\syr{\mbox{s\,yr$^{-1}$}}     
\def\kms{\mbox{km\,s$^{-1}$}}     
\def\gs{\mbox{g\,s$^{-1}$}}
\def\dm{\mbox{$\dot{M}$}}         
\def\msun{\mbox{$M_{\odot}$}}         
\def\my{\mbox{$M_{\odot}\,{\rm yr}^{-1}$}}
\def\rg{\mbox{$R_{\rm G}$}}         
\def\ra{\mbox{$R_{\rm A}$}}         
\def\rc{\mbox{$R_{\rm c}$}}
\def\peq{\mbox{$P_{\rm eq}$}}         
\def\ts{\mbox{$\tau_{\rm s}$}}
\def\ss{\mbox{2S 0114+650}}
%--------------------------------------------------------------------------

\title{ Could 2S 0114+650 be a magnetar?} 

\author{
X.-D. Li\altaffilmark{1,2} and E. P. J. van den Heuvel\altaffilmark{2}
}

\altaffiltext{1}{
Department of Astronomy, Nanjing University, Nanjing 210093, China}
\altaffiltext{2}{
Astronomical Institute, University of Amsterdam, Kruislaan 403,
1098 SJ Amsterdam, The Netherlands}

\begin{abstract}

We investigate the spin evolution of the binary X-ray pulsar \ss, 
which possesses the slowest known spin period of $\sim 2.7$ hours.  We argue 
that, to interpret such long spin period, the magnetic field strength of 
this pulsar must be initially $\gsim 10^{14}$ G, that is, it was born as
a magnetar. Since the pulsar currently has a normal magnetic field 
$\sim 10^{12}$ G, our results present support for magnetic field decay 
predicted by the magnetar model.

\end{abstract}


\section{Introduction}

Neutron stars are thought to be born as rapidly rotating ($\sim 10$ ms)
radio pulsars created during a type II/Ib supernova explosion involving a
massive star.  The dipolar magnetic fields of radio pulsars, as inferred
from their observed spin-down rates, range from $10^8$ G to $3\times
10^{13}$ G ({\cite{tml93}). However, it has been proposed that there may
exist ``magnetars" - neutron stars with magnetic field 
strengths in excess of $\sim 10^{14}$ G (\cite{td92}). These objects have
been invoked to model soft gamma-ray repeaters (SGRs) and the $6-12$ s 
anomalous X-ray pulsars (AXPs) (\cite{td96}; see also  \cite{k93}, 1994; 
\cite{p95}; \cite{c95}; \cite{vfk95}; \cite{vg97}), though
unambiguous evidence for the existence of magnetars comes from recent 
observations of the AXP-like object 1E 1841$-$045 (\cite{vg97}), 
SGRs 1806$-$20 (\cite{k98a}) and 1900+14 (\cite{k98b}).

According to Thompson \& Duncan (1992, 1996),
magnetars are neutron stars born with millisecond periods that generate
magnetic fields above $10^{14}$ G by dynamo action due to convective
turbulence, magnetic field decay powers the X-ray and particle emission
of magnetars. It is conceivable that when the magnetic fields in magnetars
have decayed to, say, $10^{12}$ G, and accretion occurs, they are not 
distinguished from X-ray pulsars born with normal ($10^{12}-10^{13}$ G) 
magnetic fields, except that they may have relatively longer spin 
periods.  One may then expect to find magnetar descendants among 
long-period binary X-ray pulsars, because they are much more luminous in
X-rays than isolated objects accreting from interstellar medium.
Here we present arguments indicating that \ss, the X-ray pulsar with the 
slowest known spin, may have a magnetar evolutionary history.

\section{The slowest X-ray pulsar \ss}

The X-ray source \ss\ was discovered in 1977 by the SAS 3 galactic survey
(\cite{dk77}).  Its optical counterpart, LS I+65 010, was recently identified 
as a supergiant of spectral type B1 Ia (\cite{r96}).
Thus \ss\ belongs to the class of high-mass X-ray binaries (HMXBs),
systems in which a compact star - generally a neutron star - accompanies a
high-mass donor star (cf. \cite{bh91} for a review).
With a distance of 7.2 kpc derived from this spectral classification,
the X-ray luminosity is a few $10^{35}-10^{36}\,\ergs$.
An orbital period of 11.59 days was reported from optical radial velocity 
measurements by Crampton, Hutchings, \& Cowley (1985). There has been some
weak evidence of a pulsation period of $850-895$ s (\cite{y90}; \cite{k83}).
In contrast, Finley, Belloni, \& Cassinelli (1992) have discovered periodic 
X-ray outbursts with a 2.78 hour period.  The same period was confirmed by 
ROSAT observations (\cite{ftb94}).  Recent Rossi X-ray Timing Explorer (RXTE) 
observations show presence of modulation on 11.59 day orbital period as well 
as $2.7$ hour pulse period (\cite{cfp98}).
If this pulse period indeed represents the rotation period of the neutron 
star, \ss\ would be by far the slowest known X-ray pulsar.

\section{The spin evolution in \ss}

How has \ss\ been spun down to the long period ($P\simeq 10^4$ s) 
if it was formed with much shorter period ($\sim 0.01-1$ s, say)? The 
neutron star's spin evolution is divided in three phases (see \cite{dp81};
\cite{bh91}).  In the first radio pulsar phase, the star is an active 
radio pulsar and spins down by
magnetic dipole radiation. This phase ends when the ram pressure of the
ambient material overcomes the pulsar's wind pressure at the gravitational 
radius ($\rg=2GM/V^2$, where $G$ is the gravitation constant, $M$ is the 
neutron star mass, and $V$ is the relative velocity between the neutron star 
and the ambient material).
In the second propeller phase material enters the corotating
magnetosphere and is stopped at $\ra$, the Alf\'ven radius, where the
energy density in the accretion flow balances the local magnetic pressure.
Further penetration cannot occur owing to the centrifugal barrier; that is,
$\ra>\rc$, where $\rc=[GM(P/2\pi)^2]^{1/3}$ is the corotation radius. The 
star expels the material once it spins up the material to the local escape 
velocity at $\sim \ra$ (\cite{is75}). Propeller action 
continues until $\ra\simeq\rc$, when the centrifugal barrier is removed and
the spin period reaches its equilibrium value (\cite{bh91})
\be
\peq\simeq (20\,{\rm s}) B_{12}^{6/7}\dm_{15}^{-3/7}R_{6}^{18/7}
	   M_{1.4}^{-5/7},
\ee
where $B=10^{12}B_{12}$ G is the neutron star's dipolar  magnetic field 
strength, $\dm=10^{15}\dm_{15}\,\gs$ the mass accretion rate, $R=10^6 R_{6}$ 
cm the radius, and $M_{1.4}=M/1.4 \msun$ (Throughout this Letter,
we take $M_{1.4}=R_{6}=1$). 
In the following accretion phase the star becomes an X-ray pulsar,
and its spin evolution is determined by the net torque exerted on the star 
by the accretion flow.

The X-ray emission observed in \ss\ is most likely to be powered by 
accretion onto the neutron star via a stellar wind from the companion. 
With a mass-loss rate of a few $10^{-6}\,\my$ and a wind velocity of
$\sim 10^3 \,\kms$, the derived X-ray luminosity from a simple wind-fed
model is in accordance with the mean detected one (\cite{r96}).  
A stable accretion 
disk is unlikely to form around the neutron star, which would require an
extremely low ($\sim 200\,\kms$) wind velocity (\cite{w81}). 
Early two-dimensional numerical studies of Bondi-Hoyle accretion flow 
(e.g., \cite{mis87}; \cite{ft88}, 1989) demonstrated that temporary 
accretion disks with alternating sense of rotation possibly form in a wind 
accreting system. More recent high resolution three-dimensional 
numerical investigations (e.g., \cite{r92}, 1997) found the so-called
wind ``flip-flop" instability with the timescale of the order of hours. 
This is consistent with the torque fluctuations in wind-accreting X-ray 
pulsars (\cite{n89}), suggesting that the long-term averaged
angular momentum transferred to the neutron star by the accreted wind
material is very small.  We therefore conclude that the spin period of 
\ss\ has not been considerably changed during the present accretion phase. 
This means that the long spin period of \ss\ 
was actually attained in an earlier evolutionary phase before the companion 
star became a super-giant - it was spun down by the propeller mechanism
(\cite{is75})
when the companion star was on main-sequence and had a weaker wind.
The magnitude of the equilibrium period, as seen in equation (1), is 
determined by the magnetic field strength and mass accretion rate of the 
neutron star during this phase.

The magnetic field strength in \ss\ has not been measured, since no 
cyclotron features have been seen in its X-ray spectrum.  But there is 
indirect evidence indicating that its magnetic field strength is similar to 
those in other X-ray pulsars: Its spectrum shows the typical shape of the
usual X-ray pulsars having a power-law with an exponential high-energy
cutoff at $\sim$ 7 keV (\cite{y90}) or $\ge 15$ keV (\cite{k83}).  The iron 
emission line at about 6.4 keV, which is common among X-ray pulsars, was 
also discovered in \ss\ (\cite{y90}; \cite{abs91}).  
Observations of cyclotron lines in
X-ray pulsars imply surface fields of about $(0.5-5)\times 10^{12}$ G,
and it has been suggested that the cutoff seen in the power-law spectra of 
X-ray pulsars is related to the magnetic field strength of the neutron 
star (\cite{mm92})
\footnote{This relation is not quite accurate (see \cite{rpw93}), but
it may provide an order of magnitude estimate of the magnetic fields.}. 
This would imply a field 
of $\sim 10^{12}$ G for \ss, consistent with those for typical X-ray 
pulsars which show cutoff energies of $10-20$ keV. 

The mass of LS I+65 010 was estimated to be $16(\pm 5) \msun$ from 
evolutionary models (\cite{r96}).
For a typical mass-loss rate of $\sim 10^{-8}-10^{-7}\,\my$ from a 16
$\msun$ main-sequence star and a wind velocity of $\sim 10^3\,\kms$,
the accretion rate of the neutron star is
$\dm\sim 10^{13}-10^{14}\,\gs$. Inserting the values of $B$ and $\dm$ 
into equation (1), we find $\peq\sim 50-140$ s, nearly two orders of 
magnitude shorter than
the observed period. (Even if $B$ is enhanced to $10^{13}$ G, $\peq$ never 
exceeds $\sim 1000$ s.) This implies that the extremely long period 
of \ss\ cannot be reached via the propeller mechanism 
{\em if the pulsar has possessed a constant magnetic field of 
$\sim 10^{12}-10^{13}$ 
G}. One may argue that the accretion rate during the propeller phase 
could be much lower than that adopted here, because of a lower rate of mass 
loss from the companion star or a higher wind velocity. For example, if
$\dm$ ranges from $\sim 10^{9}\,\gs$ (for $B=10^{12}$ G) to $\sim 10^{11}
\,\gs$ (for $B=10^{13}$ G), the value of $\peq$ can be indeed raised to 
$\sim 10^4$ s. However, this would lead to a spin-down time
during the radio pulsar phase (\cite{dp81})
\begin{eqnarray}
\ts & \simeq & (2.5\times 10^{10}\,{\rm yr})B_{12}^{-1}\dm_{9}^{-1/2}V_8^{-1}
                   \nonumber \\
    & \simeq & (2.5\times 10^{8}\,{\rm yr})B_{13}^{-1}\dm_{11}^{-1/2}V_8^{-1}
\end{eqnarray}
(where $V=10^8 V_8$ cm\,s$^{-1}$), which is
much longer than the lifetime the companion star spends on main-sequence 
(generally $\lsim 10^7$ yr). 

There exists another possibility that the neutron star was born rotating
slowly ($P\sim 1$ s), so that it went directly to the propeller phase.
Propeller spin-down to $\peq$ takes (\cite{wr85})
\begin{eqnarray}
\ts & \simeq & (1.5\times 10^{10}\,{\rm yr})B_{12}^{-1/2}\dm_{9}^{-3/4}
	       P_{1}^{-3/4} \nonumber \\
    & \simeq & (1.5\times 10^{8}\,{\rm yr})B_{13}^{-1/2}\dm_{11}^{-3/4}
		   P_{1}^{-3/4}
\end{eqnarray}
(where $P_1=P/1$ s), which is still much longer than $10^7$ yr.

We are eventually led to the conclusion that \ss\ must initially
have had a magnetic field much stronger than its present value,
that is, it was born as a magnetar. Distinguished from neutron stars with 
normal magnetic field strengths, magnetars are radio-quiet, and the decaying 
magnetic fields power the X-ray and particle emission. For a magnetar in a 
binary system, during the early evolutionary phase, the pressure exerted by 
particle emission 
exceeds the stellar wind ram pressure at $\sim \rg$, preventing 
accretion onto the neutron star (\cite{td96}). After $t\simeq 10^4-10^5$ 
years, the star's spin period increases to 
\be
P\simeq (10\,{\rm s})\,t_4^{1/2} B_{15} R_6^2 M_{1.4}^{-1/2}
\ee
by magnetic dipole radiation (where $t_4=t/10^4$ yr), 
the particle luminosity decays, and the wind
material from the companion star begins to interact with the magnetosphere.
Again adopt a mass accretion rate of $\sim 10^{13}-10^{14}\,\gs$, 
the propeller effect, for a magnetic field of $\sim (1-4)\times 10^{14}$ 
G, can comfortably spin down the neutron star from $\sim 10$ s to 
$\sim 10^4$ s on a timescale $\lsim 10^5$ years (cf. equations [1] and [3]),
i.e., before the magnetic field decays significantly (see below).
After this long equilibrium period is reached, 
the period of the neutron star would remain close to it, due to
inefficient angular momentum transfer during the subsequent
wind-accretion process. 

Field decay in \ss\ should be slow enough to garantee the final, long 
spin period, and fast enough to allow a considerable reduction in the field 
within the lifetime of the system.  There exist several mechanisms for 
field decay in non-accreting neutron stars: Ohmic
decay, ambipolar diffusion and Hall drift (e.g., \cite{gr92}). 
Ohmic decay dominates in weakly magnetized ($B\lsim 10^{11}$ G) neutron 
stars, fields in intermediate strength ($B\sim 10^{12}-10^{13}$ G) decay
via Hall drift, and intense fields ($B\gsim 10^{14}$ G) are mainly affected
by ambipolar diffusion. For typical values of the
characteristic length scale ($10^5$ cm) of the flux loops through the outer
core and the core temperature ($10^8$ K), the field-decay timescale for
ambipolar diffusion through the solenoidal mode in a magnetar is around 
$3\times 10^5$ yr if $B\sim 10^{14}$ G (\cite{td96}).  Compared to 
the age (a few $10^6$ yr) of the optical companion of \ss, this implies
that the magnetar has had enough time for its field to decay to its 
current value (a few $10^{12}$ G). Similar conclusion can also been found
from the detailed calculations by Heyl \& Kulkarni (1998).

\section{Discussion}

SGRs, which are transient gamma-ray sources that undergo repeated 
outbursts, have been suggested to be the prototypes of magnetars.
There are four known SGRs, with two (SGRs 1806$-$20 and 0525$-$66) 
associated with young supernova remnants (\cite{kf93}; \cite{v94}).
Recently, pulsations at a period of 7.47 s and 5.16 s with a spin-down rate
of $2.6\times 10^{-3}\,\syr$ and $3.5\times 10^{-3}\,\syr$ were discovered
in the persistent X-ray flux of SGRs 1806$-$20 (\cite{k98a}) and 1900+14 
(\cite{k98b}),
respectively; the magnetic field strengths of $(2-8)\times 10^{14}$ G 
can be derived if the spin-down is due to magnetic dipole radiation. 
AXPs are another type of
high-energy sources that have recently joined this group of highly magnetized
neutron stars. They are a group of about eight pulsating X-ray sources
with periods around $6-12$ s (cf. \cite{sim98} for a review), 
characterized by steady spin-down, 
relatively low and constant X-ray luminosities of $\sim 10^{35}-10^{36}$ 
\ergs, and very soft X-ray spectra. So far no optical counterpart has
been detected. Nearly half of them are located at the centers of supernova
remnants, suggesting that they are relatively young ($\lsim 10^5$ years).
Dipole magnetic fields of $10^{14}-10^{15}$ G can also be derived from the
measured spin-down rates, if they are spinning down due to dipole radiation
torques. The observed X-ray luminosities, spin periods and spin-down rates 
in SGRs and AXPs follow naturally from the magnetar model (\cite{td96}).

A crucial test of the magnetar model should include the observational
evidence of the hypothesized magnetic field decay in magnetars.
This evidence is most likely to be found in slowly-rotating, binary (such as
\ss) and isolated (such as RX J0720.4$-$3125; \cite{h97}; \cite{hh98};
\cite{kvk98}) X-ray pulsars.  Figure 1 compares the 
magnetar candidates (SGRs 1806$-$20 and 1900+14 in filled rectangles,
and AXPs 1E 2259+586, 4U 0142+61, 1E 1048.1$-$5937 and 1E 1841$-$05 
in filled stars) with their possible X-ray pulsar descendants (\ss\ in open
triangle and other pulsars in open stars) in the magnetic field versus 
spin period diagram.  A schematic view of the magnetar field and spin 
evolution of magnetars is clearly seen.
The magnetic fields of SGRs and AXPs are derived from their observed
spin-down rates (\cite{k98a}, 1998b; \cite{sim98} and references therein)
under the assumption that the spin-down is caused by magnetic dipole
radiation. For \ss\ we take a canonical magnetic field of $3\times 10^{12}$ 
G with the possible range of  $10^{12}-10^{13}$ G. 
The dashed lines represent the relation between the magnetic field strength
and the equilibrium spin period with
various mass accretion rates ($10^{18}$, $10^{15}$ and $10^{12}\,\gs$),
from which an initial magnetic field of a pulsar can be constrained by the 
present spin period, given a reasonable estimate of the mass accretion rate
during the propeller phase. The detailed evolutionary track of a magnetar
depends on the particle luminosity, the properties of the stellar wind
from the companion, the magnetic field and its decay timescale.

\begin{figure}[htbp]
\centerline{\psfig{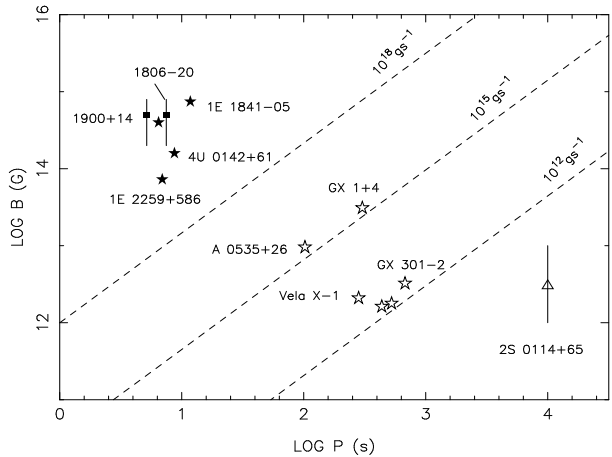}}
\caption{
The magnetic field versus spin period diagram for the candidates of 
magnetars (SGRs in filled rectangles and AXPs in filled stars) and the slow 
X-ray pulsars (in open stars except \ss\ in open triangle) with known 
magnetic fields.  The dashed lines denote the relation between the magnetic
field and the equilibrium spin period with various mass accretion rates.}
\end{figure}

In Fig. 1 we have also plotted the slow (spin periods $\gsim 100$ s) X-ray 
pulsars (A 0535+26, Vela X$-$1, GX 1+4, 4U 1907+09, 4U 1538$-$52 and 
GX 301$-$2)
with known magnetic field strengths as possible descendants of magnetars. 
Their magnetic field strengths are generally estimated from the observed 
cyclotron line features in X-ray spectra (\cite{mm98} and references therein).
For GX 1+4, it is determined from the observational evidence of the
propeller effect (\cite{c97}). These sources may have had a similar 
evolutionary history as \ss, but for them the evidence is not as unequivocal
as for \ss. Their slow spins could be accounted for in terms of the 
propeller effects with normal magnetic field strengths (e.g., \cite{is75};
\cite{wk89}, see also the dashed lines).
However, a magnetar model presents an alternative explanation that can not
be ruled out.  Actually, high magnetic field strengths ($\gsim 10^{13}$ G) 
have been measured in A 0535+26 and GX 1+4. Statistically, if the 
birth rate of magnetars is of the order of one per millennium (\cite{k98a}), 
and the X-ray lifetime of HMXBs lasts up to $\sim 10^5$ yr, there may
exist in the Galaxy $\sim 10$ binary X-ray pulsars originating from 
magnetars.

\acknowledgements
The authors thank Jan van Paradijs and the referee for helpful comments.
This work was in part supported by National Natural Science Foundation of
China and by the Netherlands Organization for Scientific Research (NWO)
through Spinoza Grant 08-0 to E. P. J. van den Heuvel.

%\newpage
%{\bf Figure Captions}

%Fig. 1 The magnetic field versus spin period diagram for the candidates of 
%magnetars (SGRs in filled rectangles and AXPs in filled stars) and the slow 
%X-ray pulsars (in open stars except \ss\ in open triangle) with known 
%magnetic fields.  The dashed lines denote the relation between the magnetic
%field and the equilibrium spin period with various mass accretion rates.

\end{document}